\pdfoutput=1

\documentclass{article}%
\usepackage{graphicx}
\usepackage{amsmath}
\usepackage{amsfonts}
\usepackage{amssymb}%
\providecommand{\U}[1]{\protect\rule{.1in}{.1in}}

\begin{document}

\author{Antony Valentini\\Augustus College}

\begin{center}
{\LARGE Beyond the Quantum}

\bigskip

\bigskip

Antony Valentini

\bigskip

\textit{Theoretical Physics Group, Blackett Laboratory, Imperial College
London, Prince Consort Road, London SW7 2AZ, United Kingdom.}

email: a.valentini@imperial.ac.uk

\bigskip

\bigskip
\end{center}

At the 1927 Solvay conference, three different theories of quantum mechanics
were presented; however, the physicists present failed to reach a consensus.
Today, many fundamental questions about quantum physics remain unanswered. One
of the theories presented at the conference was Louis de Broglie's pilot-wave
dynamics. This work was subsequently neglected in historical accounts;
however, recent studies of de Broglie's original idea have rediscovered a
powerful and original theory. In de Broglie's theory, quantum theory emerges
as a special subset of a wider physics, which allows non-local signals and
violation of the uncertainty principle. Experimental evidence for this new
physics might be found in the cosmological-microwave-background anisotropies
and with the detection of relic particles with exotic new properties predicted
by the theory.

\bigskip

\bigskip

1 Introduction

2 A tower of Babel

3 Pilot-wave dynamics

4 The renaissance of de Broglie's theory

5 What if pilot-wave theory is right?

6 The new physics of quantum non-equilibrium

7 The quantum conspiracy

\bigskip

\bigskip

\bigskip

\bigskip

\bigskip

\bigskip

Published in: \textit{Physics World}, November 2009, pp. 32--37.

\bigskip

\bigskip

\bigskip

\bigskip

\bigskip

\bigskip

\bigskip

\bigskip

\bigskip

\bigskip

\bigskip

\bigskip

\bigskip

\bigskip

\bigskip

\bigskip

\section{Introduction}

After some 80 years, the meaning of quantum theory remains as controversial as
ever. The theory, as presented in textbooks, involves a human observer
performing experiments with microscopic quantum systems using macroscopic
classical apparatus. The quantum system is described by a wavefunction -- a
mathematical object that is used to calculate probabilities but which gives no
clear description of the state of reality of a single system. In contrast, the
observer and apparatus are described classically and are assumed to have
definite states of reality. For example, a pointer on a measuring device will
show a particular reading, or a particle detector will `fire'\ a definite
number of times. Quantum systems seem to inhabit a fuzzy, indefinite realm,
while our everyday macroscopic world does not, even though the latter is
ultimately built from the former.

Quantum theory is formulated as though there is a sharply defined boundary
between the quantum and classical domains. But classical physics is only an
approximation. Strictly speaking, the classical domain does not even exist.
How does everyday reality emerge from the `unreal'\ quantum domain? What
happens to real macroscopic states as we move to smaller scales? In
particular, at what point does macroscopic reality give way to microscopic
fuzziness? What is really happening inside an atom? Despite the astonishing
progress made in high-energy physics and in cosmology since the Second World
War, today there is no definite answer to these simple questions. Standard
quantum mechanics is successful for practical purposes, but it remains
fundamentally ill-defined.

The quantum theory described in textbooks -- with an ambiguous boundary
between the quantum and classical domains -- is known as the `Copenhagen
interpretation', named after Niels Bohr's influential institute in the Danish
capital. For much of the 20th century there was a broad consensus that matters
of interpretation had been clarified by Bohr and Werner Heisenberg in 1927,
and that, despite its apparent peculiarities, the Copenhagen interpretation
should simply be accepted. But in the face of the above ambiguities, over the
last 30 years or so, that consensus has evaporated and physicists find
themselves faced with a plethora of alternative -- and radically divergent --
interpretations of their most fundamental theory.

Today, some physicists (following in the footsteps of Louis de Broglie in the
1920s and of David Bohm in the 1950s) claim that the wavefunction must be
supplemented by `hidden variables'\ -- variables that would completely specify
the real state of a quantum system. Others claim that the wavefunction alone
should be regarded as a real object, and that when the wavefunction spreads
out (as waves tend to do) this means that the system evolves into distinct,
parallel copies. The latter view -- proposed by Hugh Everett at Princeton
University in 1957 -- is particularly popular in quantum cosmology: the
wavefunction of the universe describes an ever-expanding collection of `many
worlds'. Other theorists, starting with Philip Pearle from Hamilton College in
the US in the 1970s, posit a `collapse'\ mechanism (perhaps induced by
gravity) that makes all but one part of the wavefunction disappear. And some
continue to maintain that Bohr and Heisenberg were somehow right after all.

Remarkably, today's multiplicity of viewpoints is more or less comparable to
how things were at the theory's inception in 1927 (`many worlds'\ being the
main new interpretation since then). In retrospect, the attention given to the
`Copenhagen camp'\ obscured other points of view, which never went away
entirely and which were eventually revived and have become widely known. In
particular, at the crucial 1927 Solvay conference in Brussels, no less than
three quite distinct theories of quantum physics were presented and discussed
on an equal footing: de Broglie's pilot-wave theory, Schr\"{o}dinger's wave
mechanics, and Born and Heisenberg's quantum mechanics.%

\begin{figure}
[ptb]
\begin{center}
\includegraphics[
natheight=6.503900in,
natwidth=6.511400in,
height=2.9547in,
width=2.9581in
]%
{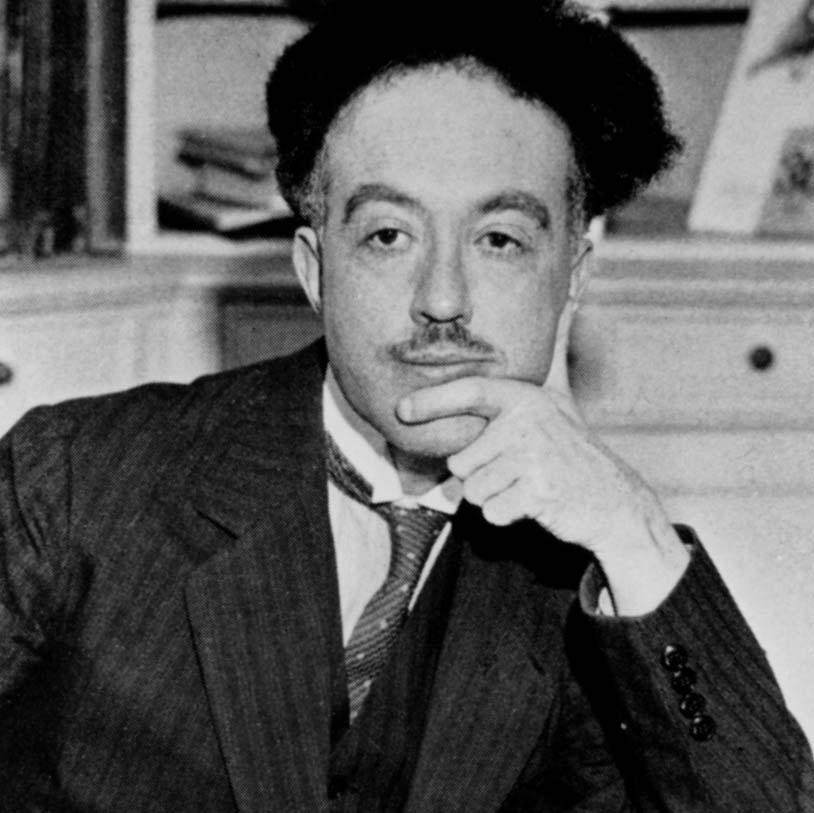}%
\caption{Louis de Broglie presented his pilot-wave theory at the 1927 Solvay
conference in Brussels. Among those present were Einstein, Dirac,
Schr\"{o}dinger, Bohr and Pauli.}%
\end{center}
\end{figure}

According to de Broglie's theory, particles (such as electrons) are point-like
objects with continuous trajectories `guided'\ or choreographed by the
wavefunction. From a modern point of view, we would say that the trajectories
are `hidden variables'\ (because their precise details cannot be seen at the
present time). Schr\"{o}dinger, in sharp contrast, presented a theory in which
particles are localized wave packets moving in space that are built entirely
out of the wavefunction -- a view that is reminiscent of modern theories of
wave packet `collapse'\ (though Schr\"{o}dinger did not propose a collapse
mechanism). According to Born and Heisenberg, neither picture is correct, and
the idea of definite states of reality at the quantum level cannot be
maintained in a way that is independent of human observation.

Standard historical accounts are, however, misleading. They say little about
de Broglie's theory or about the extensive discussions of it that took place
at the 1927 Solvay conference; and the little that is said is mostly mistaken.
In effect, de Broglie's theory was essentially written out of the standard
history of quantum physics.

It has taken some 80 years for de Broglie's theory to be rediscovered,
extended and fully understood. Today we realize that de Broglie's original
theory contains within it a new and much wider physics, of which ordinary
quantum theory is merely a special case -- a radically new physics that might
perhaps be within our grasp.

\section{A tower of Babel}

The `great quantum muddle'\ can be traced back to 1927 when, contrary to
folklore, the participants at the fifth Solvay conference distinctly failed to
arrive at a consensus (this is clear from the actual proceedings of the
conference). The sheer extent of the disagreement among the participants was
captured in a perceptive gesture by Paul Ehrenfest, who, during one of the
discussions, wrote the following quotation from the book of Genesis on the
blackboard: `And they said one to another: Go to, let us build us a tower,
whose top may reach unto heaven; and let us make us a name. And the Lord said:
Go to, let us go down, and there confound their language, that they may not
understand one another's speech.'

Like the builders of the Tower of Babel, it was as if the distinguished
physicists gathered in Brussels could no longer understand one another's speech.

However, until recently our knowledge of what happened at the conference and
in its aftermath came entirely from accounts given by Bohr, Heisenberg and
Ehrenfest -- accounts that essentially ignore the extensive formal discussions
in the published proceedings. Particularly influential was Bohr's famous 1949
essay `Discussion with Einstein on epistemological problems in atomic
physics', published in a \textit{Festschrift} for Einstein's 70th birthday and
containing Bohr's account of his discussions with Einstein at the fifth and
sixth Solvay conferences -- discussions that, according to Bohr, centred on
the validity of Heisenberg's uncertainty principle (which prevents
simultaneous measurements of position and momentum). However, not a word of
such discussions appears in the published proceedings, in which Bohr and
Einstein are in fact relatively silent. The famous exchanges between Bohr and
Einstein were informal discussions, mainly over breakfast and dinner, and were
overheard by only a few of the other participants.

De Broglie's pilot-wave theory has been particularly neglected, and its high
profile at the conference severely downplayed. According to Max Jammer's
classic historical study \textit{The Philosophy of Quantum Mechanics}, at the
conference, de Broglie's theory `was hardly discussed at all'\ and `the only
serious reaction came from Pauli', a view that is typical of standard
historical accounts throughout the 20th century. And yet, the published
proceedings show that de Broglie's theory was in fact discussed extensively:
at the end of de Broglie's talk, there are nine pages of discussion about his
theory; while of the 42 pages of general discussion (which took place at the
end of the conference), 15 pages include discussion of de Broglie's theory.
And there were serious reactions and comments from Born, Brillouin, Einstein,
Kramers, Lorentz, Schr\"{o}dinger and others, as well as from Pauli. What
exactly was the theory that de Broglie presented?

\section{Pilot-wave dynamics}

In his report -- entitled `The new dynamics of quanta'\ -- de Broglie
presented a new form of dynamics for a many-body system. In his theory,
particle motions are determined by the wavefunction, which de Broglie called a
`pilot wave'. This function obeys the usual quantum wave equation (the
Schr\"{o}dinger equation). For a many-body system, the pilot wave propagates
in a multidimensional `configuration space', which is constructed from the
co-ordinates of all the particles involved. While it was not fully appreciated
at the time, de Broglie's pilot wave is a radically new kind of causal agent
that is more abstract than conventional forces or fields in 3D space.

De Broglie's law of motion for particles is very simple. At any time, the
momentum is perpendicular to the wave crests (or lines of constant phase), and
is proportionally larger if the wave crests are closer together.
Mathematically, the momentum of a particle is given by the gradient (with
respect to that particle's co-ordinates) of the phase of the total
wavefunction. This is a law of motion for \textit{velocity}, quite unlike
Newton's law of motion for acceleration.

De Broglie had in fact first proposed this law in 1923, for the case of one
particle. His motivation had been to arrive at a unified dynamics of particles
and waves. Experiments had demonstrated the diffraction of X-rays, from which
de Broglie deduced that photons do not always move in a straight line in empty
space. He saw this as a failure of Newton's first law, and concluded that a
new form of dynamics had to be constructed.

On the basis of his new law of motion, which he applied to material particles
as well as to photons, it was de Broglie who first predicted that electrons
would undergo diffraction. This remarkable prediction was spectacularly
confirmed four years later by Clinton Davisson and Lester Germer of Bell Labs
in their experiments on the scattering of electrons by crystals. Indeed, de
Broglie won the 1929 Nobel Prize for Physics `for his discovery of the wave
nature of electrons'.

De Broglie's earlier work -- as presented in his doctoral thesis of 1924 --
had in fact been the starting point for Schr\"{o}dinger, who in 1926 found the
correct wave equation for de Broglie's waves. In the meantime, de Broglie had
sought to derive his law of motion from a deeper theory. But by 1927 he
contented himself with proposing his pilot-wave dynamics as a provisional
measure (much as Newton had regarded his theory of gravitational
action-at-a-distance as provisional).

De Broglie showed how to apply his dynamics to explain simple quantum
phenomena. But many details and applications were missing. In particular, de
Broglie seems not to have recognized that his dynamics was irreducibly
non-local. Nor was this recognized by anyone else at the conference. The
action of the wave in multidimensional configuration space is such that a
local operation on one particle can have an instantaneous effect on the
motions of other (distant) particles.

While de Broglie had (with some help from L\'{e}on Brillouin) replied to
almost all of the many queries raised in Brussels about his theory, around
1928 he became dissatisfied. In particular, he did not understand how to give
a general account of a measurement in quantum theory. To do so requires that
the dynamics be applied to the measurement process, by treating the system and
apparatus together as one larger system. This point was not fully appreciated
until the work of Bohm in 1952. Furthermore, de Broglie was uneasy with having
a wave in configuration space that affected the motion of an individual
system. Even so, he remained sceptical of the Copenhagen interpretation.

\section{The renaissance of de Broglie's theory}

De Broglie's pilot-wave theory was resurrected in 1952 when Bohm used it to
describe a general quantum measurement (for example of the energy of an atom).
Bohm showed that the statistical results obtained would be the same as in
conventional quantum theory -- \textit{if} we assume that the initial
positions of all the particles involved (making up both `system'\ and
`apparatus') have a Born-rule distribution, that is, a distribution
proportional to the squared-amplitude of the wavefunction (as appears in
conventional quantum theory).

In pilot-wave theory, the outcome of a single quantum experiment is in
principle determined by the precise (`hidden variable') positions of all the
particles involved. If the experiment is repeated many times, then the
outcomes have a statistical spread caused by the spread in the initial
distribution of particle positions.

Furthermore, Bohm noticed that the theory is non-local: the outcome of a
quantum measurement on one particle can depend instantaneously on macroscopic
operations performed on a distant particle -- the so-called `spooky'\ action-at-a-distance.

This feature caught the attention of the Northern Irish theoretical physicist
John Bell, who devoted several chapters to pilot-wave theory in his remarkably
clear and perceptive 1987 book \textit{Speakable and Unspeakable in Quantum
Mechanics}. Here was a formulation of quantum mechanics that gave a precise,
unified description of the microscopic and macroscopic worlds, in which
systems, apparatus and observers were treated (in principle) on an equal footing.

But the theory was blatantly non-local. As is well known, in 1964 Bell showed
that certain quantum correlations required any hidden-variables theory to be
non-local (on some reasonable assumptions). For several decades this was
widely seen as a blow to the hidden-variables approach, as many physicists
thought that non-locality was unacceptable. Today, however, it is increasingly
recognized that (leaving aside the many-worlds interpretation) quantum theory
itself is non-local -- as Bell had taken pains to emphasize. Non-locality
seems to be a feature of the world, and it is a virtue of pilot-wave theory to
provide a clear account of it.

Bell made it clear that the pilot wave is a `real objective field'\ in
configuration space, and not merely a mathematical object or probability wave.
Recent work at the University of Florence by Alberto Montina (now at Canada's
Perimeter Institute for Theoretical Physics) suggests that any reasonable
(deterministic) hidden-variables theory must contain at least as many
continuous degrees of freedom as are contained in the wavefunction -- and
therefore in this sense cannot be `simpler'\ than pilot-wave theory.

\section{What if pilot-wave theory is right?}

Today, we still do not know what the correct interpretation of quantum theory
is. It is therefore important to keep an open mind, and to explore the various
alternatives. What if de Broglie's pilot-wave dynamics is a correct (or at
least approximately correct) description of nature? Here, too, there have been
misunderstandings. It is usually thought that we would have to accept that the
details of the particle trajectories can never be measured, and that non-local
actions can never be controlled. This belief is based on the fact that, with
an initial Born-rule distribution of particle positions, measurements are in
practice limited by the uncertainty principle. Many scientists rightly feel
unable to accept a theory the details of which can never be checked experimentally.

However, the correct conclusion to draw is that quantum theory is merely a
special case of a much wider physics -- a physics in which non-local
(superluminal) signalling is possible, and in which the uncertainty principle
can be violated. And furthermore, the theory itself points naturally to where
this new physics might be found. Recall that pilot-wave theory gives the same
observable results as conventional quantum theory \textit{if} the initial
particle positions have a standard Born-rule distribution. But there is
nothing in de Broglie's dynamics that requires this assumption to be made. A
postulate about initial conditions can have no axiomatic status in a theory of dynamics.

An analogy with classical physics is helpful here. For a box of gas, there is
no reason to think that the molecules \textit{must} be distributed uniformly
within the box with a thermal spread in their speeds. That would amount to
restricting classical physics to thermal equilibrium, when in fact classical
physics is a much wider theory. Similarly, in pilot-wave theory, the `quantum
equilibrium'\ distribution -- with particle positions distributed according to
the Born rule -- is only a special case. In principle, the theory allows other
`quantum non-equilibrium'\ distributions, for which the statistical
predictions of quantum theory are violated -- just as, for a classical box of
gas out of thermal equilibrium, predictions for pressure fluctuations will
differ from the thermal case. Quantum equilibrium has the same status in
pilot-wave dynamics as thermal equilibrium has in classical dynamics.
Equilibrium is a mere contingency, not a law.

\section{The new physics of quantum non-equilibrium}

We have said that pilot-wave theory contains action-at-a-distance. In
particular, the outcome of a quantum measurement on one particle can depend on
macroscopic operations performed on a distant particle. This occurs,
specifically, when the wavefunction of the particles is `entangled'. In
equilibrium, this non-local effect averages to zero and no signal can be sent
in practice. But this `cancellation'\ is merely a feature of the equilibrium
state. It is \textit{not} a fundamental feature of the world.

An analogy with coins is helpful here. Consider a box containing a large
number of coins, each one showing either heads or tails. Imagine that someone
far away claps their hands, and that through some `spooky
action-at-a-distance'\ each coin is instantly flipped over. If the coins
initially had an even ratio of heads to tails, then after the flip the ratio
of heads to tails would still be even. At the statistical level, the spooky
flip would not be noticeable. But if instead the coins started with a
`non-equilibrium'\ distribution -- say 10\% heads and 90\% tails -- then the
effect of the flip would be statistically noticeable, because afterwards there
would be 90\% heads and 10\% tails.%

\begin{figure}
[ptb]
\begin{center}
\includegraphics[
natheight=2.415900in,
natwidth=6.455800in,
height=1.5575in,
width=4.117in
]%
{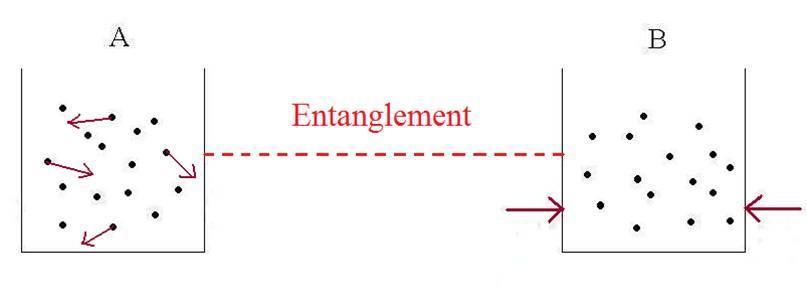}%
\caption{Two entangled boxes of particles. A local action at B -- such as
moving the walls of the box -- induces an instantaneous change in the particle
motions at A, thereby generally changing the distribution at A. For the
special case of an equilibrium distribution, the effects at A average to
zero.}%
\end{center}
\end{figure}

Something similar happens in pilot-wave theory for pairs of entangled
particles, as illustrated in figure 2. A local action at B causes an
instantaneous response in the motion of each individual particle at A. As a
result, the distribution of particle positions at A generally changes --
except in the special case of equilibrium, for which there is no net change
(at the statistical level).

Thus, if we had a large collection of non-equilibrium particles, then we could
use them for practical signalling at speeds faster than the speed of light.
Such signals could be used to synchronize clocks -- there would be an absolute
simultaneity. In the pilot-wave theory of high-energy physics, relativity
theory emerges only in the equilibrium state where such signals vanish.

It may also be shown that non-equilibrium particles could be used to perform
`subquantum'\ measurements on ordinary (equilibrium) particles -- measurements
that would violate the uncertainty principle and allow us to measure a
trajectory without disturbing the wavefunction. Essentially, the absence of
quantum noise in our `probe particles'\ would enable the experimentalist to
circumvent quantum noise in the particles being probed. Such measurements
would result in violations of standard quantum constraints, such as those on
which the security of quantum cryptography rests.

But to perform these remarkable new operations requires in the first place
that we find non-equilibrium systems. Where could these be found?

An atom in the laboratory, for example, has a past that stretches back to the
formation of stars or even earlier, during which time the atom interacted
extensively with other systems. This basic cosmological fact offers a natural
explanation for the statistical noise found in quantum systems. Indeed, there
has been ample opportunity for microscopic systems to relax to the quantum
equilibrium state, as illustrated in figure 3. In other words, given the basic
facts of astrophysics and cosmology, on the basis of de Broglie's pilot-wave
theory one would expect to find the quantum noise that we do indeed see all
around us.%

\begin{figure}
[ptb]
\begin{center}
\includegraphics[
natheight=11.647900in,
natwidth=9.383900in,
height=5.8522in,
width=4.7198in
]%
{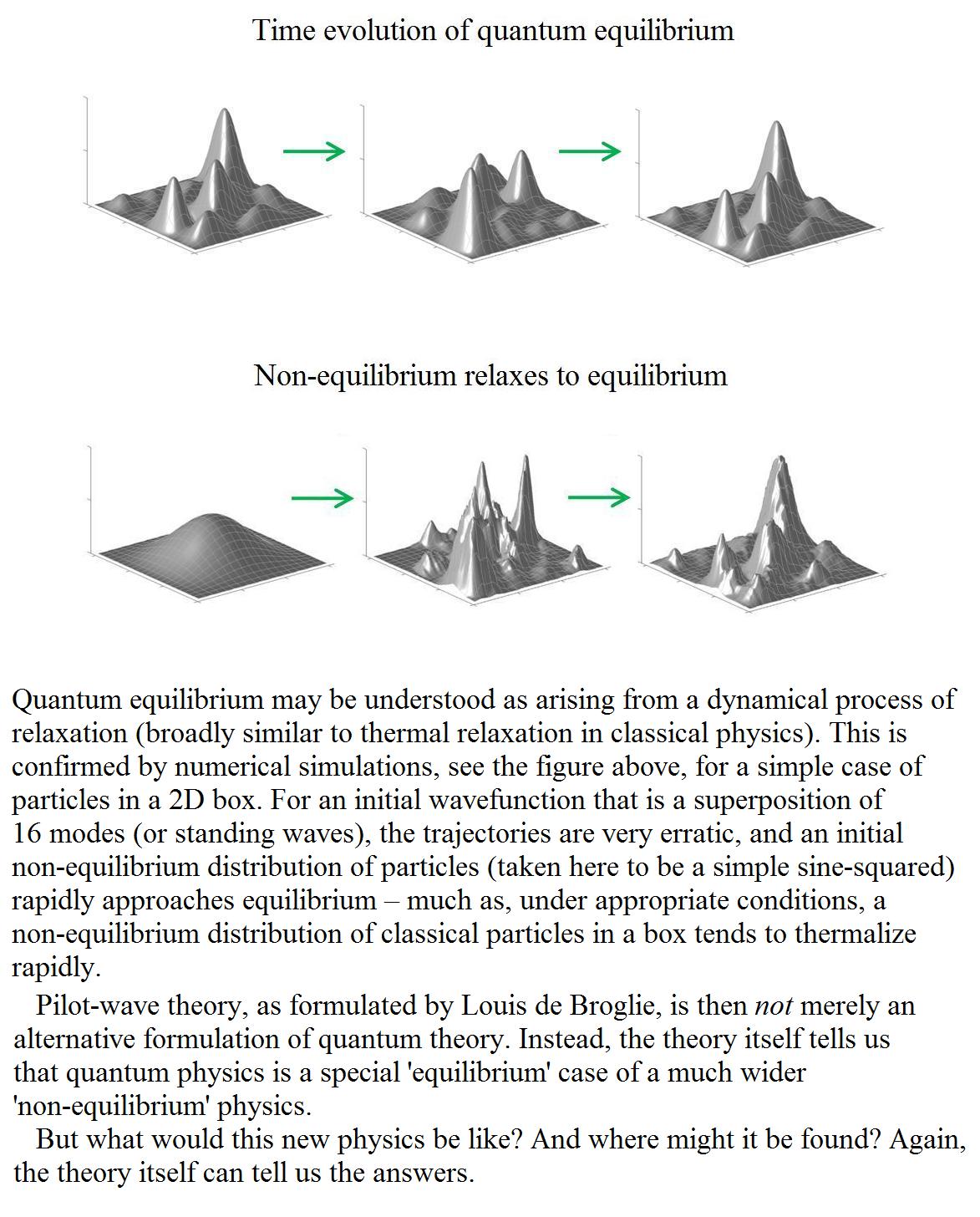}%
\caption{Relaxation and quantum equilibrium.}%
\end{center}
\end{figure}

Returning to the analogy with the box of coins, it is as if the box has been
violently shaken for a long time, so that the coins have long ago reached the
`equilibrium'\ state of an even ratio of heads to tails. And furthermore, all
the boxes of coins we have access to have undergone such long and violent shaking.

It seems natural to assume that the universe began in a non-equilibrium state,
with relaxation to quantum equilibrium taking place during the violence of the
Big Bang. On this view, quantum noise is a remnant of the Big Bang -- that is,
part of the cosmological `fossil record', rather like the cosmic microwave
background (CMB) that also pervades our universe today.

The crucial question is whether the early non-equilibrium state could have
left traces or remnants that are observable today. Given the efficient
relaxation seen in figure 3, it might be thought that any initial
non-equilibrium would quickly relax and disappear without trace. However, the
simulation shown in the figure is for a static space--time background. In the
early universe, in contrast, we must take into account the fact that space
expanded rapidly. In 2008 I showed that this can cause the initial quantum
non-equilibrium to be `frozen'\ at very large wavelengths (where, roughly
speaking, the de Broglie velocities are too small for relaxation to occur).
This result makes it possible to derive quantitative predictions for
deviations from quantum theory, in the context of a given cosmological model.

Detailed predictions remain to be worked out, but there are two obvious
avenues to explore. First, in the context of inflationary cosmology, quantum
non-equilibrium at the onset of inflation would modify the spectrum for the
CMB sky -- the hot and cold spots shown in figure 4. In other words,
measurements of the CMB can test for the presence of quantum non-equilibrium
during the inflationary phase. A second, more exciting possibility is that
some exotic particles in the very early universe stopped interacting with
other particles before they had enough time to reach equilibrium. Such
`relic'\ particles might still exist today. If we could find them, they would
violate the usual rules of quantum mechanics. (On the analogy with the boxes
of coins, some special boxes might have been shaken for a time so short that
the even ratio of heads to tails was not reached.)%

\begin{figure}
[ptb]
\begin{center}
\includegraphics[
natheight=5.280200in,
natwidth=8.640100in,
height=2.6683in,
width=4.3495in
]%
{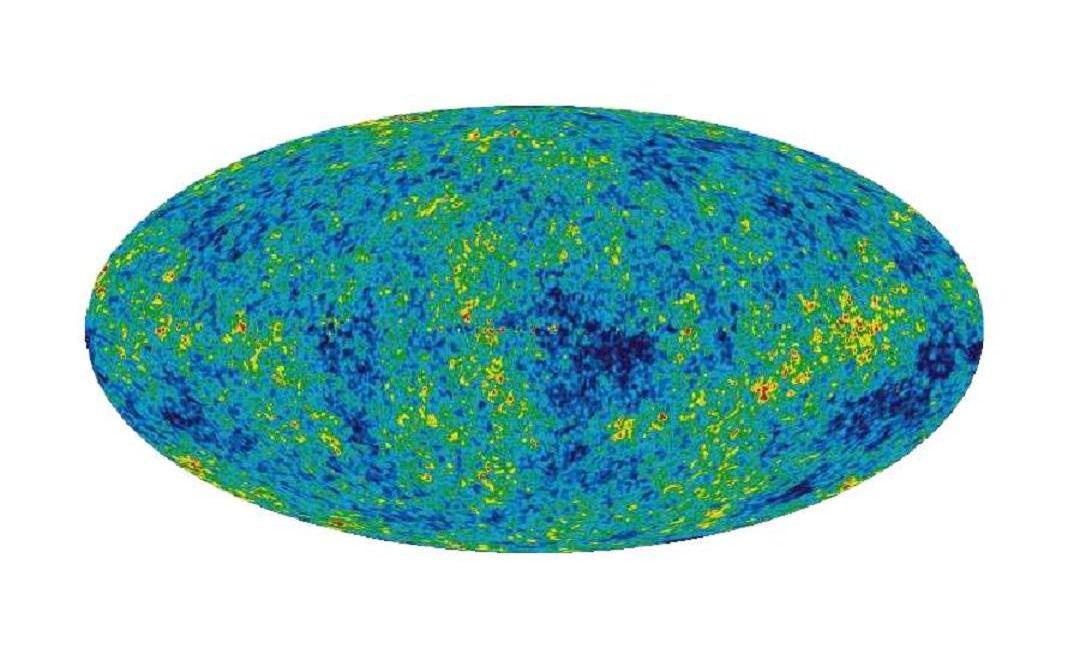}%
\caption{An all-sky map of cosmic-microwave-background (CMB) temperature
anisotropies, as measured by the Wilkinson Microwave Anisotropy Probe (WMAP)
satellite. Measurements of the CMB can be used to set limits on violations of
quantum theory in the very early universe.}%
\end{center}
\end{figure}

\section{The quantum conspiracy}

Our view of de Broglie's theory provides a very novel perspective, according
to which our local and indeterministic quantum physics emerged via relaxation
processes out of a fundamentally non-local and deterministic physics -- a
physics the details of which are now screened off by the all-pervading
statistical noise. As equilibrium was approached, the possibility of
superluminal signalling faded away and statistical uncertainty took over. Key
features of what we regard as the laws of physics -- locality, uncertainty and
the principles of relativity theory -- are merely features of our current
state and not fundamental features of the world.

But is there any independent evidence that we are confined to a special
statistical state? Arguably there is. Modern physics seems to contain a
`conspiracy'\ that prevents non-local quantum effects from being used to send
a signal. Why should non-locality be hidden in this way? The conspiracy may be
explained as a peculiarity of equilibrium, in which non-local effects are
washed out -- or averaged to zero -- by statistical noise. Out of equilibrium,
the non-locality becomes controllable and the `conspiracy'\ disappears.

To put this in perspective, recall that in the late 19th century some
theorists were concerned about the universal `thermodynamic heat death'. In
the far future, the stars would eventually burn out and all systems would
reach thermal equilibrium with each other, after which all significant
activity would cease. In such a world, in the absence of temperature
differences it would be impossible to convert heat into work -- a limitation
that would be a contingency of the state and not a law of physics. If de
Broglie's dynamics is correct, then a subquantum analogue of the classical
heat death has in fact \textit{already occurred} in our universe, presumably
some time in the remote past. In this special state, it is impossible to
convert entanglement into a non-local signal -- a limitation that is again a
contingency of the state and not a law of physics.

The slow and intermittent development of pilot-wave theory is reminiscent of
the development of the kinetic theory of gases. The work of Daniel Bernoulli
in the 18th century, and of John Waterston and others in the early 19th
century, was mostly ignored until the ideas were taken up by Rudolf Clausius
in 1857. It took decades of further work, by Maxwell, Boltzmann, Gibbs and
Einstein, among others, for the theory to yield the observable prediction of
Brownian motion. The extent to which history will repeat itself remains to be seen.

\begin{center}
{\large Further Reading}
\end{center}

G Bacciagaluppi and A Valentini 2009 \textit{Quantum Theory at the Crossroads:
Reconsidering the 1927 Solvay Conference }(Cambridge University Press) [quant-ph/0609184]

P Pearle and A Valentini 2006 Quantum mechanics: generalizations
\textit{Encyclopaedia of Mathematical Physics }(ed) J-P Fran\c{c}oise, G Naber
and T S Tsun (Amsterdam, Elsevier) pp. 265--276 [quant-ph/0506115]

A Valentini 1991 Signal-locality, uncertainty, and the subquantum H-theorem: I
and II. \textit{Phys. Lett. A} \textbf{156} 5--11; \textbf{158} 1--8

A Valentini 2007 Astrophysical and cosmological tests of quantum theory
\textit{J. Phys. A: Math. Theor.} \textbf{40} 3285--3303 [hep-th/0610032]

A Valentini and H Westman 2005 Dynamical origin of quantum probabilities
\textit{Proc. R. Soc. A} \textbf{461} 253--272 [quant-ph/0403034]

\end{document}